\documentclass[a4paper,11pt,final]{article}

\usepackage[english]{babel}
\usepackage[utf8]{inputenc}
\usepackage[T1]{fontenc}
\usepackage[top=2cm, bottom=2cm, left=2cm, right=2cm]{geometry}
\usepackage[bookmarks=false]{hyperref}
\hypersetup{
    bookmarksnumbered=true, bookmarksopen=true,
	unicode=true, colorlinks=true, linktoc=all,
	linkcolor=blue, citecolor=blue, filecolor=blue, urlcolor=blue,
	pdfstartview=FitH
}

\usepackage{amssymb}
\usepackage[square,numbers]{natbib}
\usepackage{booktabs}

\usepackage[figuresright]{rotating}
\usepackage{authblk}

\providecommand{\keywords}[1]
{
  \small	
  \textbf{\textit{Keywords---}} #1
}

\title{Link Prediction in Bipartite Networks}

\author[1]{Şükrü Demir İnan Özer} 
\author[1]{Günce Keziban Orman}
\author[2]{Vincent Labatut}

\affil[1]{Galatasaray University, Computer Engineering Department, Istanbul, Turkey}
\affil[2]{Laboratoire Informatique d’Avignon -– UPR 4128, Avignon Université, Avignon, France}

\date{28th International Conference on Knowledge-Based and \\ Intelligent Information \& Engineering Systems -- KES 2024}

\begin{document}
\maketitle

\begin{abstract}
Bipartite networks serve as highly suitable models to represent systems involving interactions between two distinct types of entities, such as online dating platforms, job search services, or e-commerce websites. These models can be leveraged to tackle a number of tasks, including link prediction among the most useful ones, especially to design recommendation systems. However, if this task has garnered much interest when conducted on unipartite (i.e. standard) networks, it is far from being the case for bipartite ones. 
In this study, we address this gap by performing an experimental comparison of 19 link prediction methods able to handle bipartite graphs. Some come directly from the literature, and some are adapted by us from techniques originally designed for unipartite networks. We also propose to repurpose recommendation systems based on graph convolutional networks (GCN) as a novel link prediction solution for bipartite networks. To conduct our experiments, we constitute a benchmark of 3 real-world bipartite network datasets with various topologies. Our results indicate that GCN-based personalized recommendation systems, which have received significant attention in recent years, can produce successful results for link prediction in bipartite networks. Furthermore, purely heuristic metrics that do not rely on any learning process, like the Structural Perturbation Method (SPM), can also achieve success.
\end{abstract}


\section{Introduction}
\label{sec:introduction}
Networks are a very common paradigm to model real-world systems, as they allow explicitly representing the interactions between the objects that constitute these systems~\cite{lfcosta-network-applications}. A whole subclass of such systems involves two different types of interacting objects, such as in user--item purchase interactions or job seeker--job advertisement applications. Their graph-based representation takes the form of \textit{bipartite} networks, which contain two distinct types of nodes, and in which links always connect nodes of different types~\cite{graph-theory-book}. Due to their more constrained structure, bipartite networks require specific processing compared to \textit{unipartite} (i.e. standard, a.k.a. homogeneous) networks. 

This is notably the case when performing link prediction, a task that consists in determining which links miss from a network known to be incomplete, or which links are likely to form in the forthcoming state of an evolving network. This is a major research topic of network science, which has been studied widely on unipartite networks~\cite{review-2011}. It is useful in a number of applications, including recommender systems, spam detection, privacy control, network routing, and bibliometrics~\cite{review-2020}. One can broadly distinguish three main approaches: (i) ranking candidate pairs of nodes using a heuristic metric that does not rely on any learning process~\cite{review-2011}, (ii) leveraging the network topological features to train a predictor into predicting such score~\cite{topologic-supervised-2006}, and (iii) perform representation learning to produce graph embeddings and use them during classification instead of features~\cite{met:n2v}.

Although extensively studied in unipartite networks, link prediction in bipartite networks remains relatively underexplored~\cite{linkpred-Kunegis, bip_lp_2018}. 
In this work, our goal is to address this gap, by proposing a comprehensive experimental study that compares a selection of link prediction methods on a bipartite network benchmark. We apply the most widespread methods from each one of the three families of approaches identified above for unipartite networks. In the process, we adapt them to bipartite networks when needed. In addition, we propose a fourth approach by leveraging state-of-the-art GCN-based personalized recommendation techniques~\cite{gnn-recsys-survey} to perform link prediction in bipartite networks. Our contributions are as follows:
\begin{itemize}
    \item Proposing a taxonomy of link prediction methods specifically designed for bipartite networks. We supplement it with visual representation to aid readers in understanding how these methods differ.
    \item Repurposing personalized recommendation models for accurate link prediction in bipartite networks.
    \item Investigating the effectiveness of using node centrality measurements as features to train supervised learning models for link prediction.
    \item Assessing the performance of 19 link prediction methods on three real-world datasets with varying sizes and densities.
\end{itemize}

The rest of this paper is organized as follows. In Section~\ref{sec:method}, we define the relevant concepts and describe in further detail the methods that we use. Then, in Section~\ref{sec:experimentandresults}, we present our experimental setup, interpret our results, and discuss their implications. Lastly, we summarize our study and findings to conclude in Section~\ref{sec:conclusion}, and discuss the future directions that could be pursued.

\section{Methods}
\label{sec:method}
Bipartite networks are characterized by their nodes being distributed over two mutually exclusive sets $U$ and $V$, and their edges $L \subseteq U \times V$ connecting only nodes from two different sets. Let us denote $L' = (U \times V) \setminus L$ as the set of links missing from the network. The problem of link prediction can then be defined as finding the proper links in $L'$ that are more likely to occur in the next evolution of the network. This prediction can be performed through many different approaches, many of which rely on the correct scoring of candidate links of $L'$ by learning-based methods. They all use the current network structure, which is encoded in $L$. In this section, we describe the different approaches that we later assess for the link prediction task in bipartite networks. For the remainder of the paper, we will refer to the nodes in $U$ as ``left nodes'' and the others in $V$ as ``right nodes''.


\subsection{Traditional Link Scores}
Traditional link scores are the metrics that quantify each link $ \in L'$ by using the structure of $L$. These scores are then ranked to choose the most likely links for predictions. To evaluate the performances of different link scores and to compare them with other approaches, we adopt a standard supervised method~\cite{eval-metrics}. We split $L$ into two sets: $L_{train}$ and $L_{test}$. We use the graph structure of $L_{train}$ to calculate these scores for all possible links, i.e. the set of $L_{test} \cup L'$. The links in this set are then ranked in decreasing order according to the scores they received. The first $|L_{test}|$ links having the highest scores constitute our prediction set, $L_{pred}$. The success of predictions is evaluated by comparing the links from  $L_{pred}$ with the links from $L_{test}$.

Most of the existing link scores rely on the concept of triadic closure observed in social networks~\cite{tcp-opsahl}, which proposes that the more neighbors two nodes share, the more likely they are to be connected. In bipartite networks, however, two nodes of the same type cannot be directly connected, and there is consequently no triangle.  Hence, we employ link prediction methods that do not rely on triadic closure. In the following scores, $x,v \in U$, $y,u \in V$,  $k_{.}$, $A$ and $l$ are  the nodes from two sets, the degree of a node, the adjacency matrix and the length of the path, respectively.

\citet{met:l3} introduced the \textit{L3 Index} after observing that the triadic closure principle was ineffective for certain networks. L3 is the degree-normalized count of paths of length three connecting two nodes.
\begin{equation}
\label{eq:l3}
    L3(x,y)= \sum_{u,v} \frac{A_{xu} A_{uv} A_{vy}}{\sqrt{k_u \times k_v}}.
\end{equation}
To compare how the length of the paths affect the link prediction performance, we also propose to use two variations of L3, that we refer to as L5 and L7. Those indices are computed the same way, except they count the paths of length five and seven, respectively.

Similarly, the \textit{Katz Index} measures the connectivity between nodes by considering the number of paths penalized with an attenuation factor $\alpha$~\cite{met:katz}.
Longer paths are penalized in an exponentially increasing manner.
\begin{equation}
\label{eq:katz}
    Katz(x,y)= \sum_{l=1}^{\infty} \alpha^l (A^l)_{xy},
\end{equation}

\textit{Local Paths} (LP)~\cite{met:lp} could be viewed as a combination of Katz and L3. It is computed like Katz, but only for paths with length of two and three. The paths of length three have an attenuation factor $\epsilon$: 
\begin{equation}
\label{eq:lp}
    LP(x,y) = (A^2)_{x,y} + \epsilon (A^3)_{x,y}.
\end{equation}
In bipartite networks, even powers of $A$ allow counting paths that connect nodes of the same type. However, in our case, $x$ and $y$ belong to different node sets, therefore the term $(A^2)_{x,y}$ from Eq.~\ref{eq:lp} takes $0$ for all $(x,y)$. 

The \textit{Preferential Attachment Index} (PA) assumes that nodes with higher degrees are more likely to be linked together~\cite{met:pa}. 
\begin{equation}
\label{eq:pa}
    PA(x,y) = k_x \times k_y.
\end{equation}

The \textit{Geodesic Distance Index} (Dist) is the reciprocal of the distance between two nodes, i.e. of the length of the shortest path between these nodes, without any normalization. By taking the reciprocal, we penalize links connecting nodes that have longer paths between them.

The \textit{Structural Perturbation Method} (SPM) studies the change in Eigenvalues when the adjacency matrix is perturbed, to measure the structural consistency of the network~\cite{spm}. The links to be predicted are ranked according to their scores in the perturbed matrix $\tilde{A}$, which is obtained through first-order perturbation.
\begin{equation}
\label{eq:spm}
    \tilde{A} = \sum^N_{k=1} (\lambda_k + \Delta \lambda_k) x_k x^T_k,
\end{equation}
where $\lambda_k$ and $x_k$ are respectively the Eigenvalue and the corresponding Eigenvector of matrix $A^R$, after the links in the perturbation set are removed from the network. Expressions $\lambda_k + \Delta \lambda_k$ and $x_k + \Delta x_k$ are the corrected Eigenvalue and its corresponding Eigenvector.

\subsection{Link Prediction via Topological Features}
\label{sec:LinkPredTopoFeat}
\begin{figure}[htb!]
  \centering
  \includegraphics[width=1\linewidth]{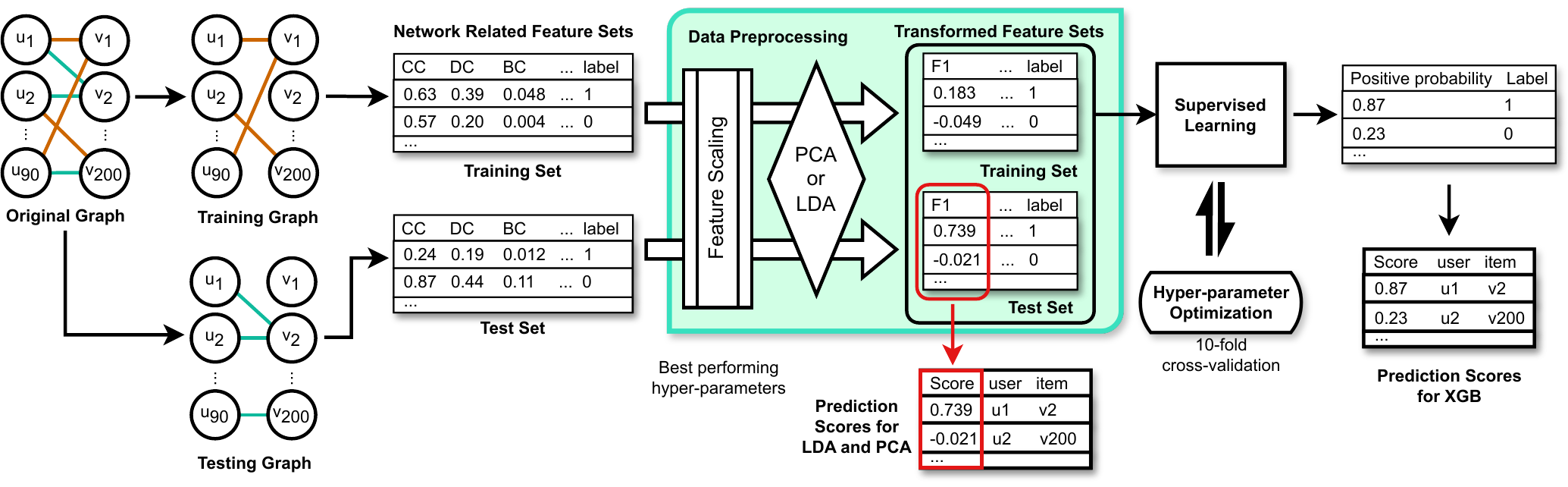}
  \caption{Representation of the link prediction method that relies on network topological features (Section~\ref{sec:LinkPredTopoFeat}).}
  \label{fig:topo}
\end{figure}

One of the common methods of link prediction relies on supervised learning techniques. In this work, we propose to represent each node pair with network-related topological features and label them with the positive or negative classes depending on whether there is a link in $L_{train}$ or not, respectively. Then, we apply classification methods. A detailed schema of the framework we propose can be seen in Fig.~\ref{fig:topo}. At first, we split the network set into training and testing subsets. Then, all node pairs from each network are represented by the PageRank~\cite{pagerank} scores of the nodes involved, as well as the PA score of each link. We also employ degree, closeness, betweenness, Eigenvector, and Katz centrality measures as reviewed by \citet{centrality-review} with detailed formulas provided in their study. 


For an unbiased training process, we use uniformly random negative sampling to include as many negative instances as positive ones, since the class numbers are highly imbalanced. We employ the XGBoost~\cite{met:xgboost} classifier model for supervised learning. It is an ensemble learning model that trains multiple gradient-boosted decision trees as estimators. We refer to this method as XGB in the rest of the paper. We also perform dimension reduction into one single feature by applying Principal Component Analysis (PCA) and Linear Discriminant Analysis (LDA). We evaluate XGB, PCA, and LDA results separately in the experiments.


\subsection{Embedding-based Link Prediction}
\label{sec:embed-based}

In recent years, network representation learning, a.k.a. graph embedding, has gained high attention because it allows for the representation of network objects, i.e., nodes, links, or whole networks, as lower-dimensional real-valued vectors. The aim of these embedding techniques is to reflect the topological differences between the objects in the network into the novel embedded features. Some popular methods developed for this goal include DeepWalk~\cite{met:dw}, node2vec~\cite{met:n2v}, GraphSAGE~\cite{met:graphsage}, and LINE~\cite{met:line}. These generic embedding methods are all unsupervised. They employ techniques such as random walks, skip-gram models, and neural networks to generate vector representations. 
However, they are not developed to handle bipartite networks and may overlook the differences in node types. Embedding techniques developed for bipartite networks are rare: in this article, we use metapath2vec, BiNE, and BiGI.

\citet{met_mp2v} introduced metapath2vec (MP2V) as an unsupervised embedding method for multipartite networks. 
Building on node2vec~\cite{met:n2v}, it applies skip-gram maximization along with a novel approach that employs meta-path guided random walk. These strategies allow capturing both structural and semantic correlations across different node types.
BiNE (Bipartite Network Embedding) attempts to capture the structural properties of bipartite networks by executing biased random walks~\cite{met_bine}. This bias allows BiNE to model the implicit and explicit relationships between the nodes of the network to preserve said structural properties such as the distribution of node degrees. 
The major limitation of BiNE is that it may not be effective for nodes with limited or no links, as it solely relies on the information obtained from observed links~\cite{met_bine}.
To overcome the drawback of random walk methods in capturing global structure,~\citet{met_bigi} proposed BiGI (Bipartite Graph Embedding via Mutual Information Maximization). This method uses two-layer multilayer perceptrons for its loss function. 

As the methods explained above take inspiration from previous node embedding algorithms such as node2vec, they also inherit their generic nature. This certainly allows for a more versatile algorithm that can cater to a wide array of tasks. At the same time, it hinders the accuracy of the algorithm, as generic approaches are unable to model the fine details that are often crucial.

\begin{figure}[htb!]
  \centering
  \includegraphics[width=0.85\linewidth]{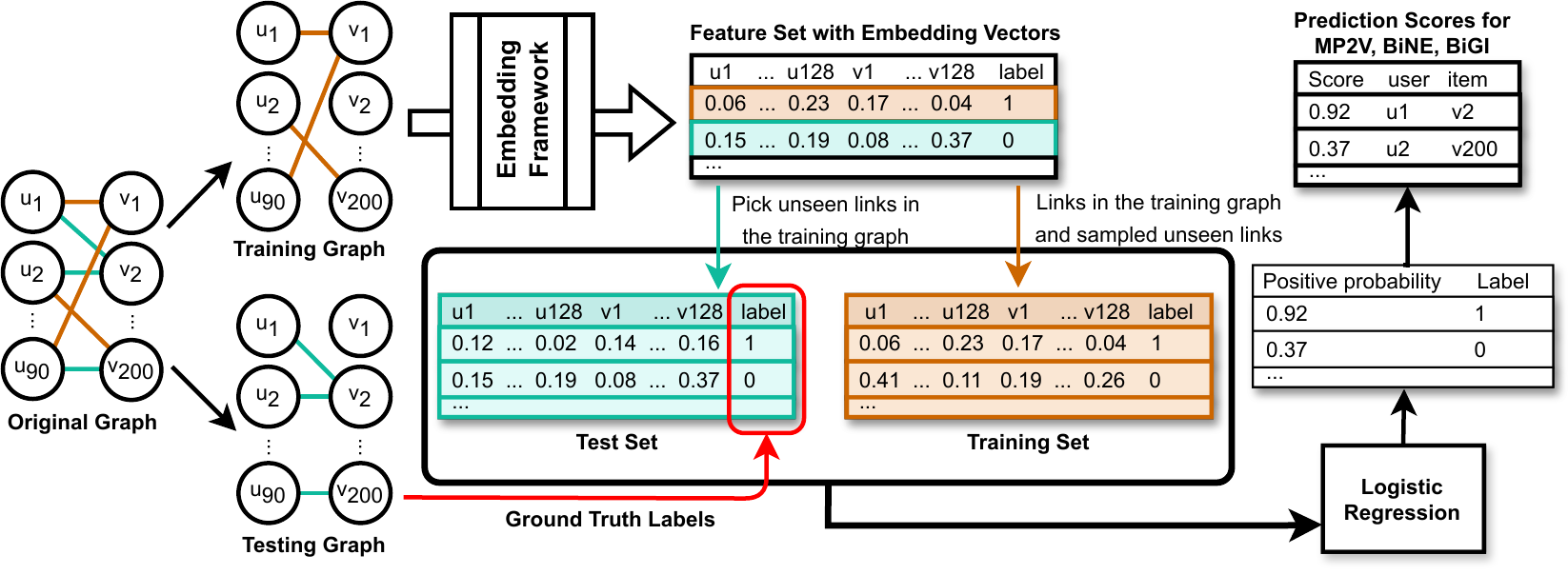}
  \caption{Representation of the link prediction method that relies on graph embeddings (Section~\ref{sec:embed-based}).}
  \label{fig:embed}
\end{figure}

These embedding methods are not directly developed for link prediction tasks. However, the openly available versions of BiNE and BiGI allow for link prediction by feeding embedding vectors to a logistic regressor while metapath2vec only produces vector embeddings of the nodes. Using the same approach used in BiNE, we utilize embeddings produced by metapath2vec to train a logistic regression model. Each node pair is represented by combining left node and right node embedding vectors. This binary classifier not only labels the node pairs but also generates a prediction score for each of them. This score estimates how likely that node pair is to belong to the positive class, which is to say, how likely it is for a link to exist between them. We treat these scores as link prediction scores and rank them accordingly, as illustrated in Fig.~\ref{fig:embed}.

\subsection{Personalized Recommendation Conversion}
\label{sec:perso-recom-conv}
Graph neural networks (GNN) and graph convolution networks (GCN) have been frequently used to handle bipartite networks in recent years. These techniques are usually employed for recommendation tasks. Basically, the user-item purchase relation is modeled under the bipartite network structure. Then, the main objective of those methods is to rank the $k$ most likely items for each user.

These approaches also generate separate embedding vectors for the nodes in $U$ and $V$, corresponding to user and item sets, respectively. 
Unlike the previously mentioned embedding-based methods however, those embedding vectors are generated specifically for the personalized recommendation task and trained with supervised learning. The embedding vectors are extracted through a GCN (or GNN), where each node's embedding vector affects those of its neighbors. On top of that, the training and prediction phases of these models are designed with this task in mind, allowing them to optimize the model for a more accurate ranking of items.

\begin{figure}[htb!]
  \centering
  \includegraphics[width=0.70\linewidth]{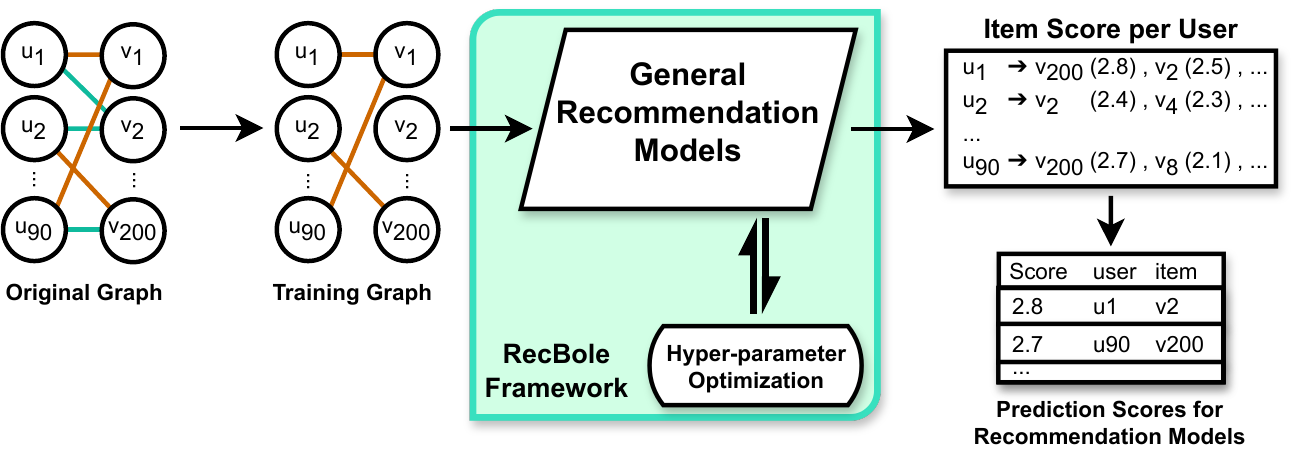}
  \caption{Representation of the link prediction method based on the repurposing of item recommendation techniques (Section~\ref{sec:perso-recom-conv}).}
  \label{fig:recbo}
\end{figure}

The output of personalized recommendation models is a ranked list of items for each user. These items are sorted based on ranking scores, which are derived from the inner product of user and item embedding vectors. In this work, we reformulate the output sets of these approaches to employ them in link prediction tasks. In order to capitalize on the effective item recommendations made from these methods, we consider the ranking score of an item $v$ for a user $u$ as the predictive score assigned to the connection between nodes $u$ and $v$, as both user and item representations influence the ranking score of the item. Our proposed approach is shown in Fig.~\ref{fig:recbo}. 

In our experiments, we use the following personalized recommendation models:
\begin{itemize}
    \item Bayesian Personalized Ranking (BPR) uses batches that consist of one user and two items as pairwise training data and optimizes a personalized ranking-focused metric to correctly rank item pairs for each user instead of scoring a single item~\cite{rec:bpr}.
    \item Neural Graph Collaborative Filtering (NGCF) utilizes high-order connectivity in user-item interactions to improve the user and item embeddings~\cite{rec:ngcf}. The initial embedding is enhanced through first-order and high-order propagation. This allows for $n$-hop neighbors to contribute to the embedding of users and items because the propagation layers can capture the collaborative signal when multiple of them are stacked.
    \item Light Graph Convolution Network (LightGCN) attempts to create node embeddings by iteratively performing graph convolution~\cite{rec:lightgcn}. Its authors suggest that feature transformation and nonlinear activation of GCNs are not necessary for the collaborative filtering task. Instead, LightGCN continuously performs weighted sum neighbor aggregation.
    \item  Self-supervised Graph Learning (SGL) aims to improve the recommendation performance of the previous GCN models like LightGCN by generating multiple representation of nodes~\cite{rec:sgl}. Different representations are obtained through modifying the network structure via node dropout, link dropout, and random walk. 
    \item Diffusion Recommender Model (DiffRec) utilizes neural networks for de-noising purposes and obtain interaction probabilities~\cite{rec:diffrec}.
\end{itemize}


\section{Experiments and Results}
\label{sec:experimentandresults}
In our experiments, we use three bipartite networks coming from different domains. \textit{MovieLens} contains users' ratings of movies on a scale of 1 to 5. We do not consider the ratings and only work with the interaction network. 
\textit{Setur}, supplied by the Turkish travel agency Setur, contains user-hotel interactions. 
\textit{LastFM} dataset contains the interactions of users with songs. As weighted links are not supported by some of the methods, the repeating interactions between the same user and song are ignored. We ignore all node and link attributes, besides node types. We apply a minimum degree filter to \textit{Setur} and \textit{LastFM} that removes left nodes (user nodes) with fewer than 8 interactions, for the sake of supervised learning experiments and prediction tasks. This filter is not applied to \textit{MovieLens}, as the lowest degree observed for the left nodes is already 20. The topological properties of the networks used in the study are given in Table~\ref{tab:dataprop}.

\begin{table}[h]
    \caption{Properties of the datasets used in this study. $\langle k \rangle$ denotes the average degree.}
    \label{tab:dataprop}
    \begin{tabular*}{\hsize}{@{\extracolsep{\fill}}lrrrrr@{}}
        \hline
        Dataset & \#Left nodes & \#Right nodes & \#Links & $\langle k \rangle$ & Density\\
        \hline
        \textit{Setur} & 324 & 1,305 & 3,798 & 4.18 & 0.00898 \\ 
        \textit{MovieLens} &943 & 1,682 & 100,000 & 68.88 & 0.06305 \\
        \textit{LastFM} & 512 & 4,014 & 18,492 & 5.91 & 0.00900 \\
        \hline
    \end{tabular*}
\end{table}



\subsection{Experimental Setup, Performance Evaluation and Hyper-parameter Tuning}
We split the network into training and test networks with 90\%  and 10\%  of the existing links, respectively. Instead of uniformly randomly selecting 10\% of all the links, we draw 10\% of the links connected to each left node. Thus, in both the training set and the test set, none of the left nodes are isolates. As a result, during prediction, we are able to rule out the well-defined cold-start problem in recommender systems, which could create additional difficulties for any of our methods.
We assess prediction quality using two key metrics: \textit{Area Under the Precision-Recall Curve} (AUPR) and \textit{ Area Under the Receiver Operating Characteristic Curve} (AUROC). The AUPR metric gauges the area under the curve plotted with precision on the $y$-axis and recall on the $x$-axis. Precision denotes the ratio of true positives to the items predicted positively, while recall is the proportion of true positives among all positive items. This metric essentially evaluates how effectively the predictor maintains high precision while lowering the positive class threshold to include more positively predicted items. Similarly, AUROC measures the area under the curve, where the $y$-axis represents recall and the $x$-axis signifies the false positive rate. The false positive rate is determined by dividing the numbers of false positives by all negative items. Both AUPR and AUROC range from 0 to 1. They do not depend on thresholds, offering general assessments that can cover different prediction scenarios. These two metrics are commonly used for link prediction performance evaluation. In~\cite{eval-metrics}, it is suggested to favor AUPR, since AUROC can be deceptive in case of class imbalance. 
It should also be noted that, in practice, AUPR takes much lower values than AUROC when assessing link prediction~\cite{eval-metrics,linkpred-Kumar}.  


For the implementation of personalized recommendation models, we use the RecBole
library, with the suggested configuration settings. 
They are tuned via random search with early stopping, using NDCG@10 as the validation metric and a training-to-validation data ratio of 8:1.
We have obtained the best performances of the learning with these parameters. The NetworkX 
Python package is used to calculate the network topological features described in Section~\ref{sec:LinkPredTopoFeat}. Link prediction indices do not require hyperparameter tuning. The same goes for PCA and LDA methods. But a random search with 10-fold cross-validation is applied to the XGB method to obtain the best performance. During the validation step, the F1 score of the positive class is used to evaluate the success of the parameters, as the prediction of negative links did not matter. Embedding-based methods are executed with the hyperparameters suggested in their respective papers to generate vector embeddings with a length of 128.

\subsection{Results}
The AUPR and AUROC results of our experiments on three different datasets are presented in Table~\ref{tab:all_res}. It is split in four-part, according to the different categories of methods introduced in Section~\ref{sec:method}. 
In addition to predictive performance, the table shows the execution time of every method on all three data sets. These do not include the time spent tuning hyper-parameters. 

For \textit{MovieLens}, SPM (category \textit{Traditional Link Scores}) gives the best results, while for \textit{Setur} and \textit{LastFM}, NGCF and DiffRec (\textit{Personalized Recommendation Conversion}) are the best, respectively. We cannot generalize this success to their whole respective categories, though, because methods from the same category exhibit a high variability in terms of performance. Still, categories \textit{Link Prediction via Topological Features} and \textit{Embedding-based Link Prediction} seem to be one step behind the others, overall. 
Independently of the method category, SPM and DiffRec seem to be the best performing methods. Especially, DiffRec obtains significantly higher AUPR scores for \textit{LastFM}. It achieves an AUPR of $0.4061$, while the highest score achieved by methods from other categories is $0.1315$, representing a $208.8\%$ increase. Dist, BiGI, and MP2V seem to get the lowest scores in general.

\begin{table}[ht]
    \caption{Performances obtained with all considered methods, for the three selected datasets, in terms of AUPR and AUROC, and runtime.}
    \label{tab:all_res}
    \resizebox{\textwidth}{!}{\begin{tabular}{@{\extracolsep{\fill}}lccrccrccrcc@{}}
        \hline
        {} & \multicolumn{3}{c}{\textit{MovieLens}} & \multicolumn{3}{c}{\textit{Setur}} & \multicolumn{3}{c}{\textit{LastFM}} \\
        Method & AUPR & AUROC & Time (s) & AUPR & AUROC & Time (s) & AUPR & AUROC & Time (s)\\
        \hline
        Katz & 0.0560 & 0.8648 & 16.51 & 0.1394 & 0.6286 & 4.00 & 0.0992 & 0.7847 & 25.27 \\
        SPM & \textbf{0.0778} & 0.8879 & 1189.00 & \textbf{0.1412} & 0.6472 & 200.00 & 0.1315 & 0.8153 & 4024.00 \\
        L3 & 0.0621 & \textbf{0.8962} & 4.92 & 0.0096 & 0.6950 & 1.35 & 0.0444 & 0.8154 & 10.60 \\
        L5 & 0.0532 & 0.8759 & 6.10 & 0.0077 & 0.6515 & 1.40 & 0.0447 & 0.7921 & 13.92 \\
        L7 & 0.0490 & 0.8648 & 7.48 & 0.0074 & 0.6379 & 1.61 & 0.0421 & 0.7921 & 12.84 \\
        PA & 0.0464 & 0.8552 & 2.20 & 0.0867 & 0.6348 & 1.24 & 0.0480 & 0.7803 & 6.24 \\
        Dist & 0.0130 & 0.7181 & 5.00 & 0.0026 & 0.6168 & 2.09 & 0.0073 & 0.7641 & 15.55 \\
        LP & 0.0561 & 0.8652 & 5.95 & 0.1398 & 0.6976 & 3.60 & 0.0993 & 0.8132 & 14.50 \\
        \hline
        XGB & 0.0179 & 0.7788 & 183.00 & 0.0039 & 0.7207 & 40.41 & 0.0337 & 0.7983 & 133.00 \\
        PCA & 0.0119 & 0.5560 & 1.92 & 0.0553 & 0.6191 & 0.32 & 0.0376 & 0.7576 & 1.47 \\
        LDA & 0.0151 & 0.7417 & \textbf{1.43} & 0.0061 & 0.6453 & \textbf{0.25} & 0.0139 & 0.7416 & \textbf{1.35} \\
        
        
        \hline
        DiffRec & 0.0613 & 0.8462 & 534.00 & 0.0541 & 0.6368 & 239.00 & \textbf{0.4061} & 0.8309 & 791.00 \\
        LightGCN & 0.0368 & 0.8266 & 184.00 & 0.1316 & 0.7126 & 13.52 & 0.0730 & 0.8342 & 9.64 \\
        BPR & 0.0270 & 0.7944 & 36.17 & 0.0473 & 0.7009 & 15.47 & 0.2152 & 0.8678 & 79.00 \\
        SGL & 0.0443 & 0.8143 & 763.00 & 0.0508 & 0.7066 & 92.00 & 0.0652 & 0.8674 & 443.00 \\
        NGCF & 0.0324 & 0.8048 & 477.00 & 0.0724 & \textbf{0.7250} & 36.31 & 0.2174 & \textbf{0.8732} & 413.00 \\
        
        \hline
        BiNE & 0.0275 & 0.8091 & {3966.00} & 0.0080 & 0.5334 & {650.00} & 0.0250 & 0.7632 & {2002.00} \\
        BiGI & 0.0368 & 0.3423 & {4834.00} & 0.0065 & 0.4434 & {204.00} & 0.0082 & 0.5379 & {1066.00} \\
        MP2V & 0.0078 & 0.5469 & {4011.00} & 0.0012 & 0.5692 & {643.00} & 0.0220 & 0.6298 & {1912.00} \\
        
        \hline
    \end{tabular}}
\end{table}

Evaluating each method individually, in the \textit{Personalized Recommendation Conversion} category, LightGCN performs best for \textit{Setur}, but receives relatively low AUPR scores for both other datasets. On top of obtaining the best AUROC scores for \textit{LastFM} and \textit{Setur}, NGCF also manages to consistently be in the top 3 when the methods are ranked according to AUPR scores. BPR, the only recommendation model that does not employ GCNs, is ranked the lowest for \textit{MovieLens} and \textit{Setur}, as expected. However, it manages to obtain the third-highest AUPR and second-highest AUROC scores for \textit{LastFM}.  
In category \textit{Traditional Link Scores}, SPM outperforms the others, most notably for \textit{LastFM}, where SPM receives a $32.43\%$ higher AUPR score compared to the second best performing method.
Following SPM, LP is consistently one of the best-performing methods. In all three datasets, there is a small difference in the AUPR scores of Katz and LP indices ($\Delta$AUPR$=0.2\%$). L5 and L7 score worse than L3 for every dataset on every metric except once, where L5 achieves a negligibly higher AUPR score than L3 for \textit{LastFM}. This shows us the ineffectiveness of using paths longer than three, since we proposed paths longer than five and seven specifically for this work. Moreover, L3 performed better than LP for \textit{MovieLens}, but worse for both other datasets. We attribute this to degree normalization in L3, as both methods are identical otherwise. Finally, Dist performed the worst. 

In category \textit{Link Prediction via Topological Features}, the highest AUPR scores are achieved by PCA. Only for \textit{MovieLens} does XGB get better results. XGB also consistently ranks highest in terms of AUROC. Finally, in category \textit{Embedding-based Link Prediction}, for \textit{MovieLens}, BiGI has a better AUPR score than others, although it has the lowest AUROC score in the entire table. For the other datasets, it cannot replicate this success and is surpassed by BiNE.

Examining the experimental results, it is evident that there is a considerable variation in the scores obtained from the performance metrics depending on the dataset. To make an interpretation of the results, we take into account the properties of the datasets seen in Table~\ref{tab:dataprop}. Indeed, as the density of the data declines, so does the accuracy of the predictions made by the L3 and Dist link prediction indices. This decrease can be explained by the difference in L3 index scores between low-probability and high-probability links becoming less pronounced.
Such a decrease, however, is not observed for the SPM, Katz, LP, and PA indices.
Nevertheless, they seem to be affected by the size of the network. \textit{Setur} and \textit{LastFM} have similar densities, but the former is smaller. This could be the reason for their success with \textit{Setur}.
Another group of methods that does not see a decline in evaluation metrics in sparse networks is \textit{Personalized Recommendation Conversion}. On the contrary, their performance seems to increase as the data becomes sparser.
This may be attributed to their unique architecture. The \textit{Personalized Recommendation} models used in our study, except for BPR, employ GCNs to generate representations of the nodes.
While metapath2vec and BiNE methods also generate node embeddings, they rely on random walks to do so. BiGI utilizes GNNs, whose effectiveness is apparent in the AUPR score achieved on the \textit{MovieLens} dataset, matching LightGCN's AUPR score.

Moreover, here, \textit{Personalized Recommendation Conversion} methods are optimized for scoring items to make accurate recommendations to users. In contrast, \textit{Embedding-based Methods} treat the link prediction task as a binary classification problem by fitting a logistic regression model to make predictions. They are similar in this way to methods that use topological network features. It is also possible that the success of these methods are limited by the simple classifier used during the prediction step. 

The methods that rely on topological features do not apply graph representational learning at all. It could be argued that this puts them at a disadvantage compared to methods that can exploit the network structure through path-based algorithms or graph representational learning. While topological features may accurately describe a node's role in the network, they do not offer any insight on how two specific nodes interact with each other. 
The same line of thought could be applied to \textit{Embedding-based} methods. They indeed generate embedding vectors, but no matter how successful this process is, those embeddings are generalized representations of nodes. By contrast, \textit{Personalized Recommendation Conversion} models generate embeddings for the explicit purpose of assigning a score to a pair of nodes. 
As a result, although we do not use exact-ranked items but convert them to link prediction, it seems to be effective in terms of AUPR and AUROC accuracy.

Evaluation metrics used to score predictions should also be considered while interpreting our results. The link prediction task primarily focuses on predicting future links, in other words, positive instances. High AUPR scores indicates that the predictor can maintain a high 
ratio of true positives to positively predicted items, as the positive class g threshold gets lower to include more positives. This is observed as high precision values while recall increases. In our case, the AUPR metric is sensitive to small changes in the number of true positives. In the link prediction task, where there is an abundance of negative instances, AUROC should be handled with care. Due to the high number of negatives, the false positive rate may not vary significantly even after substantial changes in the number of false positives. For these reasons, we primarily compare AUPR scores while paying attention to AUROC scores.


Regarding the execution time of the methods, generally, it can be seen that the more successful methods, such as SPM and GCN-based, take longer to make predictions.
Although it is not the best-performing method on any data set, LP shows considerable success in prediction accuracy while maintaining low runtime.
The quickest category is \textit{Link Prediction via Topological Features}, due to optimized algorithms of PCA and LDA. 
Except for SPM, which is an outlier, \textit{Traditional Link Scores} are the second quickest. Nevertheless, their scalability falls behind that of other methods. Indeed, on average, they take 18 times longer to finish processing \textit{LastFM} compared to \textit{Setur}. Most of the indices predict link scores through the adjacency matrix, which represents every link possible in the network. As a consequence, such methods do not scale well in large but sparse real-world networks. Other methods seem to scale off the number of links present in the network in addition to the number of nodes. This is evidenced by their lengthy execution times for \textit{MovieLens}, which is smaller than for \textit{LastFM} in terms of the number of nodes and all possible links.
\textit{Embedding-based Link Prediction} methods take longer to work than other methods, despite their low prediction accuracy. This can be attributed to the fact that they use logistic regression to obtain prediction scores, which may not be able to handle the embedding dimensions.

To summarize our findings, methods of category \textit{Personalized Recommendation Conversion}, that use a GCN architecture, are more effective in link prediction compared to unsupervised embedding methods based on random walks, traditional link scores, and classifiers trained with network topological features. We attribute the success of these models to the fact that they specifically generate node embeddings with a supervised technique using the already purchased/preferred items and go through a training phase to make recommendations, which can be formulated as link prediction problems on a per-user basis. \textit{Traditional Link Scores} suffer in sparse networks due to a lack of preexisting information. But surprisingly, SPM, a simple link score, highly performs in several cases. It outperforms all other methods we employed in our experiments except DiffRec and NGCF. 
In cases where it is vital to make predictions as fast as possible, the LP index can be used to get results in the shortest possible time without too much compromising predictive performance.

\section{Conclusion}
\label{sec:conclusion}
In this paper, we compared different link prediction methods that could be applied to bipartite networks. We evaluated 19 different methods that we classified into four different categories, with three different datasets of varying sizes and density. Most of the methods that we used were not previously applied to link prediction in bipartite networks. Hence, our work carries especially experimental novelties for link prediction methodologies in bipartite networks. The key result of our experiments can be listed as it follows.

\begin{itemize}
    \item Our major contribution as a novel link prediction approach, repurposing recommendation models that generate personalized ranked lists of items for link prediction tasks in bipartite networks, outperforms almost all the other methods.
    \item Specifically, the DiffRec model achieved an AUPR score of $0.4061$ on the \textit{LastFM} network, which has a low density ($0.009$) and a high number of nodes ($4,526$).
    \item SPM, which is a traditional link score, is one of the best performing methods. Although this method was not developed for bipartite networks, our experiments show that it can be a very good link predictor for bipartite networks as well.
    \item Methods that use bipartite network embeddings and network topological features for link prediction tasks produce the worst results.
\end{itemize}
Personalized Recommendation methods generate node embeddings within the GCN architecture using supervised learning from interacted states, unlike unsupervised embedding methods. Our experimental results indicate that for link prediction, supervised learning with GCNs yields superior performance in graph representation learning.
This study could be further expanded by including more datasets with varying network sizes, link counts, and densities. An ensemble model could be used for personalized recommendation models, since our results imply that different recommendation models were able to better capture the structural properties of different networks.

\section*{Acknowledgements}
This work is supported by the Galatasaray University Research Fund (BAP) within the scope of project number FBA-2023-1204, titled \textit{\"Oneri Sistemlerinin Performans Değerlendirmesi}, and the bilateral project of the Scientific and Technological Research Council of Türkiye (TÜBITAK), under grant number 122N701, with CampusFrance, within the scope of Hubert Curien Partnerships (PHC) project number 49032VB.

\bibliography{Ozer2024.bib}
\bibliographystyle{elsarticle-harv}

\end{document}